\documentclass[a4paper,11pt]{article}
\usepackage{fullpage}
\usepackage{amsmath, amssymb}
\usepackage{epsfig}
\def\Li{\operatorname{Li_2}}
\def\Im{\operatorname{Im}}

\def\negcdot{\negthickspace\cdot\negthinspace}
\def\negthincdot{\negthinspace\cdot\negthinspace}
\def\negtwo{\negthickspace\negthickspace}
\def\Re{\operatorname{Re}}

\begin{document}
\begin{center}
\Large{\bf{One loop scalar functions in the heavy quark effective theory}}
\vskip1cm
\large{J. Zupan} \footnote{jure.zupan@ijs.si}
\vskip0.5cm
{\it J. Stefan Institute, Jamova 39, P.O.Box 3000, SI-1001 Ljubljana, Slovenia}
\end{center}
\abstract{We find general solutions for the dimensionally regularised scalar one loop  three-point  and four-point functions with one heavy quark propagator. The  scalar one-point function vanishes, while the expression for the  two-point function has been found before.  For the latter  we give a detailed derivation. We also discuss  some  special cases and compile useful formulae.}

\section{Introduction}
Since its early applications \cite{Isgur:vq} the heavy quark symmetry has been one of the key ingredients in the theoretical investigations  of hadrons containing a heavy quark. It has been successfully applied to the heavy hadrons spectroscopy, to the inclusive as well as to a number of exclusive decays (for reviews of the heavy quark effective theory and related issues see \cite{Neubert:1993mb} or \cite{Manohar:dt}). To describe interactions with not too energetic light mesons, the heavy quark symmetry has been combined with chiral symmetries leading to the heavy hadron chiral perturbation theory (HChPT) \cite{Casalbuoni:1996pg}. 

The one loop calculations within the  heavy quark effective theory are considerably simplified if the light-quark masses are neglected. Very common in the HChPT is a similar approximation, with the finite contributions omitted, while  only the leading logs are retained \cite{Goity:1992tp},\cite{Falk:1993fr}. This approximation is, however,  becoming less adequate with the increasing precision of data coming from B factories. To go beyond the leading log approximation and/or take into account the counterterms appearing at the next order in the chiral expansion,  the general solutions for the  one loop scalar functions need to be considered. In the context of the HChPT  a general solution for the one loop scalar two-point function with one heavy quark propagator has been found in \cite{Boyd:1994pa},\cite{Stewart:1998ke}. We extend this calculation  and find solutions for the scalar  three-point  and four-point functions with one heavy quark propagator.

The vector and tensor one loop functions can then be  expressed in terms of the scalar one loop functions using the algebraic reduction  \cite{Passarino:1978jh}. Also, the  one loop scalar functions with two or  more heavy quark propagators can be  expressed in terms of the one loop scalar functions with just one heavy quark propagator. For the case of the equal heavy-quark velocities this can be accomplished using the relation
\begin{equation}
\frac{1}{v\negcdot q-\Delta}\frac{1}{v\negcdot q-\Delta'}=\frac{1}{\Delta-\Delta'}\left(\frac{1}{v \negcdot q -\Delta}-\frac{1}{v \negcdot q-\Delta'}\right).
\end{equation}
For unequal heavy quark velocities  techniques developed in \cite{Bouzas:1999ug} can be used.

The scalar  one-loop functions with heavy quark propagators can be derived also directly from the scalar functions of the full theory by using the threshold expansion \cite{Beneke:1997zp} (see also appendix B of \cite{Davydychev:2001ui}). This technique has recently been used for the calculation of the scalar and tensor three-point functions with one and two heavy quark propagators \cite{Bouzas:2001py},\cite{Bouzas:2002xi}. In the present paper, however, we will not  follow the approach of Bouzas et al. \cite{Bouzas:1999ug},\cite{Bouzas:2001py},\cite{Bouzas:2002xi} but rather do the calculation from scratch.

The paper is organized as follows: in  section 2. we make some general remarks and list useful relations used further on in the calculation. In section 3. we calculate one and two point functions. We continue with the calculation of the three-point function in section 4.  and with  the four-point function in section 5.

\section{General remarks and useful relations}
Before we start with the calculation, let us first list some useful relations and the conventions that are going to be used further on. The greater part of this section is a review of the relations and the conventions used in \cite{'tHooft:1978xw} with certain modifications.
The major difference between the conventional  one loop scalar functions  and the one loop scalar functions with one heavy quark propagator is the appearance of the  propagator linear in the integration variable $q$. Therefore  a modified version of the standard Feynman parameterization is used
\begin{equation}
\begin{split}
\frac{1}{\big(\prod_{i=1}^N A_i\big) B}=& N! \int_0^\infty\negtwo 2 d \lambda \int \prod_{i=1}^N d u_i \; \frac{\delta(1-\sum_i u_i) \;\prod_i \Theta(u_i)}{[\sum_{i=1}^N A_i u_i +2 B \lambda]^{N+1}}\\
=& N!\; \frac{ \int_0^\infty   2 d\lambda \int_0^1 dx_1 \int_0^{x_1} dx_2 \cdots \int _0^{x_{N-2}}dx_{N-1} \hfill}{[A_1(1-x_1)+ \sum_{j=2}^{N-1}A_j (x_{j-1}-x_j) + A_N x_{N-1} +2 B \lambda ]^{N+1}} , \label{eqS:24}
\end{split}
\end{equation}
where $\Theta(u)$ is the Heaviside function $\Theta(u)=1$ for $u>0$ and zero otherwise. In the calculation $A_i$ are going to be ``full'' (inverse)  propagators $((q+p_i)^2-m_i^2+i\delta)$ and $B$ the heavy quark propagator $(v\negthincdot q - \Delta + i \delta)$. Note also, that the leading power of $q^2$ in the denominator has increased from the left-hand side's $(q^2)^{N+1/2}$ to the right-hand side's $(q^2 )^{N+1}$. The integration over $q$ has  been made more convergent, but then  another integration over infinite range (integration over $\lambda$) has been introduced through the parameterization.

A very useful identity used in the calculation is
\begin{equation}
\begin{split}
\frac{1}{[(q+p_1)^2-m_1^2+i \delta][(q+p_2)^2-m_2^2+i \delta]}= &\frac{\alpha}{[(q+p_1)^2-m_1^2+i \delta][(q+l)^2-M^2+i \delta]}\\
&+\frac{1-\alpha}{[(q+p_2)^2-m_2^2+i \delta][(q+l)^2-M^2+i \delta]} \label{eqS:9},
\end{split}
\end{equation}
where $\alpha$ is an arbitrary parameter and
\begin{align}
l&=p_1+\alpha(p_2- p_1),\\
M^2&=(1-\alpha)m_1^2+\alpha m_2^2-\alpha(1-\alpha)(p_2-p_1)^2.
\end{align}
The parameter $\alpha$ can then be chosen at will. It is useful to keep it real, though. Then there are no ambiguities connected with the shift of  the integration  variable $q$, that is  performed, as usual, before the Wick rotation.  For instance $\alpha$ can be chosen such  that  $M^2=0$. If $(p_2-p_1)^2\leqslant (m_1-m_2)^2 $  or $(p_2-p_1)^2\geqslant (m_1+ m_2)^2$, then  $\alpha$ is real.  If one of the masses is made to be zero, the integration is simplified considerably (as will be seen in the calculation of the four-point function  \eqref{eqS:5}). The other option used below is to set $\alpha$ such that $l^2=0$. This can be done for real $\alpha$ if (but not only if) one of $p_1$, $p_2$ or  $p_1\pm p_2$ is timelike. This shows, that in general product of propagators at least  one internal or one external mass can always be set to zero, even with  $\alpha$ restricted to be real.

In doing the integrals the following procedure proves to be very useful. Consider
\begin{equation}
\int_0^\infty \negtwo 2 d\lambda \int_0^1 dx \frac{1}{[a x^2 + b \lambda^2 + c x \lambda+ \cdots]}.
\end{equation}
The integration over $x$ can be simplified by the change of the integration variables $\lambda=\lambda' + \beta x$, where $\beta$ is chosen such, that the coefficient in front of $x^2$ vanishes, i.e. $\beta$ has to solve the equation $b \beta^2+ c\beta + a=0$. Then the integrand is linear in $x$, so the integration over $x$ is trivial. The integration bounds are 
\begin{equation}
\begin{split}
\int_0^\infty \negtwo 2 d\lambda \int_0^1 dx \cdots &= \int_0^1 dx \int_{-\beta x}^\infty 2 d \lambda' \cdots=\\
&= \int_0^\infty \negtwo 2 d\lambda \int_0^1 d x \cdots+ \int_{-\beta}^0 2 d \lambda \int_{-\lambda/\beta}^1 dx \cdots \label{eqS:25} .
\end{split}
\end{equation}

As the results of the integration, the functions such as logarithms, dilogarithms (Spence functions) and hypergeometric functions will appear. Since the arguments of the functions will in general lie in the complex plane it is necessary to discuss the conventions used. The logarithms used in this paper have a cut along the negative real axis. For $x$ exactly  negative real we use the prescription $\ln(x)\to \ln(x+i \epsilon)$, where $\epsilon>0$ is a positive infinitesimal parameter. In other words,  $\ln(x)=\ln|x|+i\pi$ for $x$ negative real\footnote{Note that this prescription does not change the calculation of the logarithms away from the negative real axis. In particular it does not change the value of a logarithm with an argument that already has an infinitesimal but nonzero imaginary part. For more discussion on this point see text after Eq. \eqref{eqS:37}. }. In particular $\ln(-1)$ is defined to be $\ln(-1)=i\pi$. Of course this choice is completely arbitrary and at the end of the calculation one has to check that results are independent of this choice. Using this definition for the logs of the negative real arguments the logarithm of an inverse is  
\begin{equation}
\ln\left(\frac{1}{x}\right)=-\ln(x)+2 \pi i \Re^{(-)}(x), \label{eqS:55}
\end{equation}
with
\begin{equation}
\Re^{(-)}(x)=\left\{
\begin{aligned}
1&\; ;\; & x \text{ on negative real axis},\\
0&\; ; \; & \text{otherwise}.
\end{aligned}
\right. \label{eqS:47}
\end{equation}
Note that the change from the usual rule for the  logarithm of an inverse is just on the negative real axis. For the arguments away from the negative real axis the function $\Re^{(-)}(x)$ is exactly zero and everything is as usual.
The  logarithm  of a product is
\begin{equation}
\ln(a b)= \ln(a) +\ln(b) +\eta(a,b),
\label{eqS:17}
\end{equation}
where $\eta$ function is \footnote{Note, that in comparison with \cite{'tHooft:1978xw}, the $\eta$ function has been extended also to the negative real arguments (cf. discussion after Eq. \eqref{eqS:8} and Eq. \eqref{eqS:51}). For arguments away from the negative real exis (also if by an infinitesimal amount) it is the same as in \cite{'tHooft:1978xw}.}
\begin{equation}\label{eqS:50}
\eta(a,b)=\left\{
\begin{aligned}
\;&
\begin{aligned}
2 \pi i \big\{ & \Theta(-\Im a) \Theta(-\Im b)\Theta(\Im ab)\\
&-\Theta(\Im a) \Theta(\Im b)\Theta(-\Im ab)\big\} 
\end{aligned} \; ; \; &\text{  $a$ and  $b$ {\it not} negative real},\\
\;& 
- 2 \pi i \big\{  \Theta(\Im a)+\Theta(\Im b) \big\}
\;  ;\; &\text{ either $a$ or $b$ negative real},\\
\;&
-2 \pi i \;  ;\; &\text{  $a$ and  $b$ negative real}.
\end{aligned}
\right. 
\end{equation}
The normal rule for the logarithm of a product applies for these important cases
\begin{equation}
\begin{split}
\ln(ab)&=\ln(a)+\ln(b) ;\; \text{$\Im a$ and $\Im b$ have the opposite sign},\\
\ln(a/b)&=\ln(a)-\ln(b) ;\; \text{$\Im a$ and $\Im b$ have the same sign},
\end{split}\label{eqS:54}
\end{equation}
with $a$, $b$ not negative real.

\section{One- and two-point functions}
In this section we will concentrate on the calculation of the dimensionally regularized one-point and two-point functions in the heavy quark effective theory
\begin{align}
-\frac{1}{16 \pi^2} \bar{A}_0 (\Delta)&=\frac{i \mu^{\epsilon}}{(2\pi)^{n}}  \int d^{n}q \frac{1}{(v \negcdot q -\Delta+i\delta)},\\
-\frac{1}{16 \pi^2} \bar{B}_0 (m,\Delta)&=\frac{i \mu^{\epsilon}}{(2\pi)^{n}}  \int d^{n}q \frac{1}{(v \negcdot q -\Delta+i\delta)(q^2-m^2+i \delta)} , \label{eqS:2}
\end{align}
where $\delta$ is  a positive infinitesimal parameter, $n=4-\epsilon$, while $m$ and $\Delta$ are real. The general solution for the two-point function  has  been found by Stewart in Ref. \cite{Stewart:1998ke} (see also  \cite{Casalbuoni:1996pg} and references therein). In this section we will derive Stewart's result.

We start with the integral
\begin{equation}
I_r=\frac{i \mu^{\epsilon}}{(2\pi)^n }\int d^n q \frac{1}{(v\negcdot q -\Delta+i \delta)(q^2-m^2+i\delta)^r}.
\end{equation}
Using the  Feynman parameterization  \cite{Manohar:dt}
\begin{equation}
\frac{1}{a^r b^s}=2^s \frac{\Gamma(r+s)}{\Gamma(r)\Gamma(s)} \int_0^\infty \negtwo d\lambda \frac{\lambda^{s-1}}{(a + 2b \lambda)^{r+s}},
\end{equation}
we get
\begin{equation}
I_r=\mu^\epsilon \frac{(-1)^{-r}}{\Gamma(r)}\frac{2 \Gamma(r+1-\frac{n}{2})}{(4 \pi)^{\frac{n}{2}}}\int_0^\infty \negtwo d\lambda (\lambda^2+2 \lambda \Delta+m^2 -i \delta) ^{\frac{n}{2} -r -1}.
\end{equation}
 The integral can be expressed in terms of the hypergeometric function. Introducing the new variable  $\lambda'=\lambda +\Delta$ and then splitting the integration interval for  negative $\Delta$ we get
\begin{equation}
\int_\Delta^\infty\frac{d \lambda'}{({\lambda'}^2-\Delta^2+m^2 -i \delta)^N}=\int_{-|\Delta|}^{|\Delta|} \frac{d \lambda}{(\lambda^2 -\Delta^2+m^2-i\delta)^N} \Theta(-\Delta) +\int_{|\Delta|}^{\infty}\frac{d\lambda}{(\lambda^2-\Delta^2+m^2-i\delta)^N},
\end{equation}
where  we write $N=r+1-n/2$ for short. Another change of variables $u=\lambda^2$ leads to
\begin{equation}
\begin{split}
(m^2-\Delta^2-i\delta)^{-N}\Big\{&\int_{\Delta^2}^\infty \frac{du}{2 \sqrt{u}}\Big[\frac{u}{m^2-\Delta^2-i\delta}+1 \Big]^{-N}\;+ \\
\;&+ 2 \int_0^{\Delta^2} \frac{d u}{2 \sqrt{u}} \Big[\frac{u}{m^2-\Delta^2-i\delta}+1 \Big]^{-N} \Theta(-\Delta)\Big\}\label{eqS:3}.
\end{split}
\end{equation}
These integrals can be expressed in terms of the   hypergeometric functions ${}_2F_1(\alpha, \beta;\gamma; z)$ through  the identities \eqref{eqS:22}, \eqref{eqS:23} listed in the appendix \ref{app:A}. Using the transformation formula  \eqref{eqS:14} together with ${}_2F_1(0,\beta;\gamma;z)={}_2F_1(\alpha,0;\gamma;z)=1$ we arrive  at
\begin{equation}
\begin{split}
I_r=\mu^\epsilon\frac{(-1)^{-r}}{\Gamma(r)} \frac{ 2 \Gamma(N)}{(4\pi)^{\frac{n}{2}}}\bigg[ &-\Delta (m^2 -\Delta^2-i\delta)^{-N} {}_2F_1\negthickspace \left(N,\frac{1}{2};\frac{3}{2};\frac{-\Delta^2}{m^2-\Delta^2-i \delta}\right)\\
&+ \frac{\Gamma\left(N+\frac{1}{2}\right)\Gamma\left(\frac{1}{2}\right)}{(2N-1) \Gamma(N)} (m^2-\Delta^2-i \delta)^{\frac{1}{2}-N}\bigg] \label{eqS:1},
\end{split}
\end{equation}
where $N=r+1-n/2$. 

Let us first discuss the case when $r$ is equal to zero or negative integer, i.e. when the integrand, apart from the heavy quark propagator, is a polynomial. The integrals for the physical case $n\to 4$ are divergent, but we can make sense of it through analytic continuation.  At fixed  $r$ the integral $I_r$ is taken to be an analytic function of the complex dimension $n$. For $r$ equal to zero or negative integer and $n/2 \ne \mathbb{Z}$ all functions appearing in \eqref{eqS:1} are finite,  apart from $\Gamma(r)$ that is infinitely large. Thus $I_r$ vanishes for $r$ zero or negative integer everywhere in the $n$ complex plane apart from the points on the real axis with integer  $n/2 \geqslant r+1 $. Analytic continuation of $I_r$ \eqref{eqS:1} is then equal to zero  in the whole $n$ plane.  Integrals over polynomials (and one heavy quark propagator) are in the dimensional regularisation thus equal to zero. In particular, the one-point scalar function $\bar A_0=0$.

For the two-point function we have $r=1$ and therefore $N=\epsilon/2$. So the two point function is
\begin{equation}
\begin{split}
-\frac{2}{16 \pi^2} \left(4 \pi \mu^2  \right)^\frac{\epsilon}{2}\; &\Big\{\Gamma \negthickspace\left(\frac{\epsilon}{2}\right) (-\Delta) (m^2-\Delta^2-i\delta)^{-\frac{\epsilon}{2}}\; {}_2F_1\negthickspace \left(\frac{\epsilon}{2},\frac{1}{2};\frac{3}{2};\frac{- \Delta^2}{m^2-\Delta^2-i\delta}\right)\\
&- \pi (m^2-\Delta^2-i\delta)^{\frac{1}{2}}\Big\}.
\end{split}
\end{equation}
Since $\Gamma(\epsilon/2)\to 2/\epsilon -\gamma+{\cal O}(\epsilon)$ we have to expand hypergeometric function around $\epsilon/2=0$  in order to get the finite terms correctly
\begin{equation}
{}_2F_1\negthickspace\left(\frac{\epsilon}{2},\frac{1}{2};\frac{3}{2};z\right)=1+ \frac{\epsilon}{2} \frac{\partial}{\partial N}\left. {}_2F_1\negthickspace\left(N,\frac{1}{2};\frac{3}{2};z\right)\right|_{N=0}+\dots .
\end{equation}
The partial derivative can  be found using the series expansion \eqref{eqS:15} and is
\begin{equation}
\frac{\partial}{\partial N} \; \left. {}_2F_1\negthickspace\left(N,\frac{1}{2};\frac{3}{2};z\right)\right|_{N=0}=-\ln(1-z)-z^{-\frac{1}{2}} \ln\Big(\frac{1+\sqrt{z}}{1-\sqrt{z}}\Big)+2 .
\end{equation}
This leads us to the final result for the two-point function
\begin{equation}
\begin{split}
\frac{i \mu^{\epsilon}}{(2\pi)^{4-\epsilon}}& \int d^{4-\epsilon}q \frac{1}{(v \negcdot q -\Delta+i \delta)(q^2-m^2+i \delta)}=\\
& \qquad \qquad \frac{2\Delta}{(4 \pi)^2} \Big\{ \frac{2}{\epsilon}-\gamma+\ln 4\pi -\ln\Big(\frac{m^2}{\mu^2}\Big)+2 -2 F\left(\frac{m}{\Delta}\right)\Big\}\label{eqS:33} ,
\end{split}
\end{equation}
where F(x) is a function as defined in \cite{Stewart:1998ke} valid for both positive and negative $\Delta$ (while $m$ is always taken to be positive real)
\begin{equation}
F\left(\frac{1}{x}\right)=\left\{
\begin{aligned}
\;&\frac{1}{x}\sqrt{x^2-1} \; \ln(x+\sqrt{x^2-1}+i \delta)  \; ; &\;|x|>1,\\
-&\frac{1}{x}\sqrt{1-x^2}\left[\frac{\pi}{2}-\tan^{-1} \left(\frac{x}{\sqrt{1-x^2}}\right)\right]\; ;& \;|x|\le 1 ,
\end{aligned}\right.
\end{equation}
with $\delta$ an infinitesimal positive parameter. Note that for $x<-1$,  $ F(1/x)$ has an imaginary part that corresponds to the particle creation. Also, the two point function has to be a continuous function of $\Delta$, as can be  seen from \eqref{eqS:2} or \eqref{eqS:3}. It is easy to check that for $|x|=1$ the two-point function is continuous as then $F(\pm 1)=0$. The two-point function is also continuous for $\Delta\to 0$. Even though $F(1/x)$ diverges as $x\to0$, the two point function \eqref{eqS:33} is finite and equal to $m/8 \pi$.

Finally, for $r \geqslant 2$ both $\Gamma(r)$ and $\Gamma (N)$ in \eqref{eqS:1}  are finite in the limit $n\to 4$, so that  Eq. \eqref{eqS:1} can be used directly,  with $n$ set to $n=4$.

\section{Three-point scalar function}
The one loop scalar  three-point function with one heavy quark propagator is given by
\begin{equation}
-\frac{1}{16 \pi^2} \bar{C}_0 (v,k, \Delta, m_1,m_2)=\frac{i \mu^\epsilon}{(2 \pi)^{n}} \int d^{n} q \frac{1}{(v \negcdot q -\Delta+i\delta)(q^2-m_1^2 +i \delta)((q+k)^2-m_2^2+i\delta)} .\label{eqS:Three}
\end{equation}
This integral is finite in 4 dimensions, so that $\epsilon$ can be set to zero.  Using the Feynman parameterization \eqref{eqS:24} we get
\begin{equation}
\bar{C}_0=- \int_0^\infty \negtwo 2 d\lambda \int_0^1 dx \frac{1}{[\lambda^2 +k^2 x^2 +2 v\negcdot k x \lambda -(k^2+m_1^2-m_2^2)x + 2\Delta \lambda+m_1^2-i\delta]}\label{eqS:42} .
\end{equation}
The integration over $x$ can be made trivial through the change of variables $\lambda=\lambda'+\alpha x$. We choose $\alpha$ to be the solution of 
\begin{equation}
(k+\alpha v)^2=0, \label{eqS:49}
\end{equation}
 as then the term quadratic in $x$ is zero. The solution is $\alpha_{1,2}=(-v \negthincdot k\pm \sqrt{(v\negthincdot k)^2-k^2})$ and is real for any real four-vector $k^\mu$ (this can be easily seen by going in the frame, where $v^\mu=(1,0,0,0)$ with  the square root then equal to  $\sqrt{\vec{k}^2}$). Changing the integration order as in \eqref{eqS:25}, and then integrating over $x$, we get
\begin{equation}
\begin{split}
\bar{C}_0=- \bigg[ &\int_0^\infty \frac{2 d\lambda}{(A\lambda +B)}\ln\bigg(\frac{\lambda^2+C\lambda +D+(A\lambda +B)}{\lambda^2+C \lambda +D}\bigg)\\
&+\int_0^1\frac{2 \alpha d\lambda}{(-A\alpha \lambda +B)}\Big\{ \ln[\alpha^2 \lambda^2-C\alpha\lambda+D+(-A\alpha \lambda+B)]\\
&\qquad\qquad\qquad \qquad\quad-\ln [\alpha^2 \lambda^2-C\alpha \lambda +D+(-A \alpha \lambda +B)\lambda]\Big\}\bigg]\label{eqS:4} ,
\end{split}
\end{equation}
where
\begin{equation*}
\begin{matrix}
&A=2(v\negcdot k +\alpha),  &B=2 \Delta \alpha +m_2^2-m_1^2-k^2 , \\
&C=2\Delta , &D=m_1^2-i\delta .
\end{matrix}
\end{equation*}
and $\alpha$ one of the solutions $\alpha_{1,2}$ of the quadratic equation \eqref{eqS:49}.  In \eqref{eqS:4} we have used the fact that D is the only complex parameter and split the logarithm in the second integrand. The integrands of both the first and the second integral have vanishing residua.  In the second integral we then  add and subtract the values of the logarithms for $\lambda=B/A\alpha$ and write
\begin{equation}
\begin{split}
\bar{C}_0=-\frac{1}{A}\Bigg[&\int_0^\infty\negtwo\frac{2 d\lambda}{\lambda +B/A} \ln \bigg(\frac{\lambda^2+(A+C)\lambda +(B+D)}{\lambda^2+C\lambda+D}\bigg)\\
-&\int_0^1 \negtwo\frac{2 d\lambda}{\lambda-B/(A\alpha)} \Big(\ln \big[\alpha^2\lambda^2-(A+C)\alpha \lambda+(B+D)\big]\\
&\qquad\qquad \qquad\qquad\qquad\qquad-\ln\big[\alpha^2\lambda_0^2-(A+C)\alpha \lambda_0+(B+D)\big]\Big)\\
+&\int_0^1\negtwo \frac{2 d\lambda}{\lambda-B/(A\alpha)}\Big(\ln\big[\alpha(\alpha-A)\lambda^2+(B-C\alpha)\lambda +D\big]\\
&\qquad\qquad\qquad\qquad\qquad\qquad -\ln\big[\alpha(\alpha-A)\lambda_0^2+(B-C\alpha)\lambda_0+D\big]\Big)\bigg]\label{eqS:21} ,
\end{split}
\end{equation}
with $\lambda_0=B/(A\alpha)$. Note, that all three integrals in \eqref{eqS:21} have integrands with vanishing residua.  These integrals can be reduced into the  sums of the dilogarithms, where care has to be taken regarding the imaginary parts of the arguments of the logarithms. The  solutions of the integrals can be found in the appendix \ref{app:B}. The solution of the first integral can be found in  \eqref{eqS:27}, \eqref{eqS:41}, with the definitions \eqref{eqS:36}, \eqref{eqS:48}  (where $a_1=a_2=0$, note also the minus sign), while the solutions to the last two integrals can be found in \eqref{eqS:28}, \eqref{eqS:29}. Using the functions $S_3$ and $I_2$ defined in the appendix \ref{app:B} the three-point function \eqref{eqS:Three} finally reads
\begin{equation}
\begin{split}
\bar{C}_0= \frac{2}{A}\Big[\;& I_2\left(0,0,A+C,B+D,C,D,-B/A\right)\\
+&S_3\left(\alpha^2,-(A+C)\alpha,B+D,B/A\alpha\right)\\
-& S_3\left(\alpha(\alpha-A),B-C\alpha,D,B/A\alpha\right)\Big].\label{ThreePoint}
\end{split}
\end{equation}
Note that the value of the three-point function in \eqref{ThreePoint} does not depend on which of the solutions $\alpha_{1,2}$ of the equation \eqref{eqS:49} is used. This can be used as a useful check in the numerical implementation.

The solution is simplified considerably if $k^2=0$. Then the $x$ integration in \eqref{eqS:42} is trivial. Proceeding similarly as above we arrive at
\begin{equation}
\bar{C}_0(v,k,\Delta,m_1,m_2)\Big|_{k^2=0}=-\frac{1}{v \negcdot k} \sum_i \rho(\kappa_i) \left[ \Li\left(\frac{\lambda_0}{\lambda_0-\kappa_i}\right)+\frac{1}{2} \ln^2(\lambda_0-\kappa_i)\right],
\end{equation}
where $\lambda_0=(m_1^2-m_2^2)/(2 v \cdot k)$, while  $\kappa_i$ are the solutions of
\begin{equation}
\begin{split}
&\lambda^2+ 2 (v \negcdot k+\Delta)\lambda +m_2^2-i \delta=(\lambda-\kappa_1)(\lambda-\kappa_3),\\
& \lambda^2+2 \Delta \lambda +m_1^2-i \delta=(\lambda-\kappa_2)(\lambda-\kappa_4),\label{eqS:43}
\end{split}
\end{equation}
and $\rho(\kappa_i)=(-1)^{i+1}$.

The solution is even further simplified if besides $k^2=0$ also $m_1=m_2=m$. Then
\begin{equation}
\bar{C}_0(v,k,\Delta,m,m)\Big|_{k^2=0}=-\frac{1}{v \negcdot k} \sum_i \rho(\kappa_i) \left[ \frac{1}{2} \ln^2(-\kappa_i)\right],
\end{equation}
with $\kappa_i$ and $\rho(\kappa_i)$ given in \eqref{eqS:43}. The three point function in this limit has been calculated before and is given explicitly in \cite{Fajfer:2001ad} (see Eq. (A10) of \cite{Fajfer:2001ad}). The two expressions agree completely. A number of numerical checks between numerically integrated expression \eqref{eqS:4} and final expression \eqref{ThreePoint} have been performed as well.   

\section{Four-point function}
The scalar four-point function with one heavy quark propagator is defined as
\begin{equation}
\begin{split}
-\frac{1}{16 \pi^2} \bar{D}_0(v,p_1,&p_2,\Delta,m_1,m_2,m_3)=\\
&\frac{i \mu^\epsilon}{(2\pi)^{n}} \int \frac{\; \; d^{n}q \hfill }{(v\negcdot q-
\Delta)(q^2-m_1^2)((q+p_1)^2-m_2^2)((q+p_2)^2-m_3^2)} , \label{eqS:10}
\end{split}
\end{equation}
where the $i\delta$ prescription has been omitted in the notation. Again, the integral is convergent and $\epsilon$ can be set to zero. Since the Feynman parameterization \eqref{eqS:24} is not symmetric in $A_i$ and $B$,  the elegant transformation used in the calculation of the conventional four-point function \cite{'tHooft:1978xw} and further improved in \cite{Denner:qq} unfortunately cannot be applied. Instead, one repeatedly  uses the propagator identity \eqref{eqS:9} to solve the integral \eqref{eqS:10}. Since the parameter $\alpha$ in propagator identity \eqref{eqS:9} has to be real, the calculation differs depending on the values of the external momenta $p_1$ and $p_2$. 

First we take up the case, when one of the following inequalities is true $p_{1,2}^2\geqslant (m_1+m_{2,3})^2$ or $p_{1,2}^2\leqslant (m_1-m_{2,3})^2$. If necessary, we renumber the momenta and reshuffle the propagators in \eqref{eqS:10} in such a way that either $p_1^2\geqslant (m_1+m_{2})^2$ or $p_{1}^2\leqslant (m_1-m_{2})^2$ in order to simplify the discussion. Then  we use  the propagator identity \eqref{eqS:9} on the second and the third propagators of \eqref{eqS:10} 
\begin{equation}
\begin{split}
\frac{1}{(q^2-m_1^2+i \delta)((q+p_1)^2-m_2^2+i \delta)}= &\frac{1-\alpha}{[(q+p_1)^2-m_2^2+i \delta][(q+l)^2-M^2+i \delta]}\\
&+\frac{\alpha}{[q^2-m_1^2+i \delta][(q+l)^2-M^2+i \delta]},
\end{split}
\end{equation}
where $\alpha$ is an arbitrary parameter and
\begin{align}
l&=\alpha p_1,\\
M^2&=(1-\alpha)m_1^2+\alpha m_2^2-\alpha(1-\alpha)p_1^2  .
\end{align}
We choose $\alpha$ such that $M^2=0$. This  is satisfied by real  $\alpha$ if either  $p_1^2\geqslant (m_1+m_{2})^2$ or $p_{1}^2\leqslant (m_1-m_{2})^2$ as has been assumed above. The scalar four point function is then
\begin{equation}
\begin{split}
\frac{i}{(2\pi)^{4}} &\int d^{4} q\frac{1-\alpha}{(v\negcdot q -\Delta)[(q+p_1)^2-m_2^2][(q+p_2)^2-m_3^2][(q+l)^2+i\delta]}\;+\\
\frac{i }{(2\pi)^{4}} &\int d^{4} q\frac{\alpha}{(v\negcdot q -\Delta)[q^2-m_1^2][(q+p_2)^2-m_3^2][(q+l)^2+i\delta]}, \label{eqS:5}
\end{split}
\end{equation}
with $\alpha$ the solution of
\begin{equation}
p_1^2 \alpha^2 + (m_2^2-m_1^2-p_1^2)\alpha +m_1^2=0 . \label{eqS:32}
\end{equation}
To calculate the two integrals in \eqref{eqS:5} it suffices to consider 
\begin{equation}
\begin{split}
-\frac{1}{16\pi^2}\tilde{D}_0&(v,k_1,k_2,k_3,\Delta,M_1,M_2)=\\
&\frac{i}{(2\pi)^{4}}\int d^{4}q\frac{1}{(v\negcdot q-\Delta)[(q+k_1)^2-M_1^2][(q+k_2)^2-M_2^2][(q+k_3)^2+i\delta]},\label{eqS:6}
\end{split}
\end{equation}
where the $i\delta$ prescription has not been written out explicitly in the first three propagators.

Using the Feynman parameterization \eqref{eqS:24} and integrating over $q$ we arrive at
\begin{equation}
\begin{split}
\tilde{D}_0= \int_0^\infty \negtwo 2d\lambda \int_0^1 \negtwo dx \int_0^x\negtwo  dy [&p_{23}x^2+p_{12}y^2+(p_{13}-p_{23}-p_{12})xy+\lambda^2+(P_2-P_3)x\lambda+P_3\lambda \\
&+(P_1-P_2)y\lambda +(-p_{23}+M_2^2)x+(p_{23}-p_{13}+M_1^2-M_2^2)y- i \delta]^{-2} ,
\end{split}\label{eqS:52}
\end{equation}
with
\begin{equation}
\begin{split}
p_{ij}&=(k_i-k_j)^2,\\
P_i&=2(v\negcdot k_i+\Delta).\\
\end{split}\label{eqS:44}
\end{equation}
To simplify the integration we introduce new variables $y=x y'$ and $\lambda=x\lambda'$. The integration limits are then $\int_0^1 dx \int_0^x dy \int_0^\infty 2 d\lambda \to \int_0^1 dx\int_0^1 x dy' \int_0^\infty x 2 d\lambda'$. Since $x$ is positive and $\delta$ an infinitesimal parameter of which only the sign matters, the extra factor of $x^2$ in the numerator can be canceled against the similar factor in the denominator. After the cancellation the denominator is linear in $x$. The integration over $x$ is now  trivial and yields
\begin{equation}
\begin{split}
\tilde{D}_0=\int_0^\infty \negtwo 2 d\lambda \int_0^1 dy &[(p_{23}-p_{13}+M_1^2-M_2^2)y +P_3 \lambda -p_{23}+M_2^2-i\delta]^{-1} \times\\
& [p_{12}y^2+\lambda^2+(P_1-P_2) y\lambda +P_2\lambda +(-p_{12}+M_1^2-M_2^2)y+M_2^2-i\delta]^{-1} .\label{eqS:53}
\end{split}
\end{equation}
To cancel the $y^2$ term in the integral above a new variable $\lambda=\lambda'+\beta y$ is introduced, with $\beta$ chosen to solve
\begin{equation}
\beta^2+(P_1-P_2)\beta+ p_{12}=0 . \label{eqS:45}
\end{equation}
The solutions are  real for $(k_2-k_1)^\mu$ real.
We get
\begin{equation}
\tilde{D}_0=\big(\int_0^\infty \negtwo 2 d\lambda \int_0^1 \negtwo dy+\int_{-\beta}^0 \negtwo 2 d\lambda \int_{-\lambda/\beta}^1 \negtwo \negtwo dy\;\big) \frac{1}{[a_1 y+b_1 \lambda +c_1] [a_2y +b_2\lambda +c_2+d_2y\lambda +\lambda^2]} , \label{eqS:7}
\end{equation}
with
\begin{equation}
\begin{matrix}
&
\begin{aligned}
a_1&=\beta P_3+p_{23}-p_{13}+M_1^2-M_2^2,\\
b_1&=P_3,\\
c_1&=M_2^2-p_{23}-i\delta,\\
{}&{}
\end{aligned}
&\begin{aligned}
 a_2&=\beta P_2+M_1^2-M_2^2-p_{12},\\
b_2&=P_2,\\
c_2&=M_2^2-i\delta,\\
d_2&=2\beta+P_1-P_2 .
\end{aligned}
\end{matrix}\label{eqS:46}
\end{equation}
After $y$ integration we arrive at
\begin{equation}
\begin{split}
\tilde{D}_0=& \frac{1}{(\lambda_1-\lambda_2)}\frac{1}{(b_1d_2-a_1)}\Bigg[\\
&\int_0^\infty \negtwo\frac{2 d \lambda}{\lambda-\lambda_1} \bigg\{ \ln\left(\frac{b_1\lambda+c_1}{b_1\lambda+a_1+c_1}\right)-\ln\left(\frac{\lambda^2+b_2\lambda+c_2}{\lambda^2+(b_2+d_2)\lambda+a_2+c_2}\right)\bigg\}\\
-&\int_0^\infty \negtwo\frac{2 d \lambda}{\lambda-\lambda_2} \bigg\{ \ln\left(\frac{b_1\lambda+c_1}{b_1\lambda+a_1+c_1}\right)-\ln\left(\frac{\lambda^2+b_2\lambda+c_2}{\lambda^2+(b_2+d_2)\lambda+a_2+c_2}\right)\bigg\}\\
-&\int_0^1\negtwo\frac{2 d\lambda}{\lambda+\lambda_1/\beta} \bigg\{ \ln\left(\frac{(a_1-\beta b_1)\lambda+c_1}{-\beta b_1\lambda+a_1+c_1}\right) -\ln\left(\frac{\beta(\beta -d_2)\lambda^2+(a_2-b_2\beta)\lambda+c_2}{\beta^2\lambda^2-\beta (b_2+d_2)\lambda+a_2+c_2}\right)\bigg\}\\
+&\int_0^1 \negtwo\frac{2 d\lambda}{\lambda+\lambda_2/\beta} \bigg\{ \ln\left(\frac{(a_1-\beta b_1)\lambda+c_1}{-\beta b_1\lambda+a_1+c_1}\right) -\ln\left(\frac{\beta(\beta -d_2)\lambda^2+(a_2-b_2\beta)\lambda+c_2}{\beta^2\lambda^2-\beta (b_2+d_2)\lambda+a_2+c_2}\right)\bigg\}\Bigg], \label{eqS:8}
\end{split}
\end{equation}
with $\lambda_{1,2}$ the solutions of 
\begin{equation}
(b_1d_2-a_1)\lambda^2+(a_2b_1-a_1b_2+c_1d_2)\lambda+a_2c_1-a_1c_2=(b_1d_2-a_1)(\lambda-\lambda_1)(\lambda-\lambda_2).\label{lam12}
\end{equation}
Note that the integrands above have vanishing residua, i.e. arguments of the two  logarithms in the integrands are the same for $\lambda=\lambda_{1,2}$. Note as well, that the infinitesimal imaginary parts of $c_1$ and $c_2$, the $-i \delta$ in \eqref{eqS:46}, have to be equal. They originate from the same infinitesimal parameter in \eqref{eqS:52} that after the integration over $x$ appears twice in \eqref{eqS:53}. The size of the infinitesimal parts of $\lambda_{1,2}$ compared to $-i\delta$ are thus unambiguously defined and have to be kept track of until the end of the calculation.

 The integrals \eqref{eqS:8} can be expressed in terms of the dilogarithms. This has been done in the appendix \ref{app:B}. The solution for the first two integrals can be found in  \eqref{eqS:27}, with the definitions in \eqref{eqS:36}, \eqref{eqS:48}, while the solution for the  last two integrals can be found in \eqref{eqS:31}, \eqref{eqS:26}. Together with Eqs. \eqref{eqS:5}, \eqref{eqS:6} and the cascade of abbreviations \eqref{eqS:46}, \eqref{eqS:45}, \eqref{eqS:44} this gives the complete solution of the four point function with at least one external momentum $p_1$, $p_2$ satisfying $p_{1,2}^2\geqslant (m_1+m_{2,3})^2$ or $p_{1,2}^2\leqslant (m_1-m_{2,3})^2$. Collecting the terms and rearranging the last two  propagators in \eqref{eqS:10} if necessary, the four point function for  $p_{1}^2\geqslant (m_1+m_{2})^2$ or $p_{1}^2\leqslant (m_1-m_{2})^2$ finally reads
\begin{equation}
\begin{split}
\bar{D}_0(v,p_1,p_2,\Delta,m_1,m_2,m_3)=&(1-\alpha) \tilde{D}_0(v,p_1,p_2,l,\Delta,m_2,m_3)\\
&+\alpha \;\tilde{D}_0(v,0,p_2,l,\Delta,m_1,m_3),\label{eqS:Tildefun}
\end{split}
\end{equation}
with $l=\alpha p_1$ and $\alpha$ the solution of $p_1^2\alpha^2+(m_2^2-m_1^2-p_1^2)\alpha +m_1^2=0$, while
\begin{equation}
\begin{split}
\tilde{D}_0&(v, k_1,k_2,k_3,\Delta,M_1,M_2)=\frac{1}{(\lambda_1-\lambda_2)}\frac{2}{(b_1d_2-a_1)}\Big[\\
 &I_2\left(-c_1/b_1,-(a_1+c_1)/b_1,b_2,c_2,b_2+d_2,a_2+c_2,\lambda_1\right)\\
-&I_2(-c_1/b_1,-(a_1+c_1)/b_1,b_2,c_2,b_2+d_2,a_2+c_2,\lambda_2)\\
-&I_1(\beta^2-\beta d_2,\beta^2,a_2-\beta b_2,-\beta(b_2+d_2),a_1-\beta b_1,-\beta b_1,c_2,a_2+c_2,c_1,a_1+c_1,-\lambda_1/\beta)\\
+&I_1(\beta^2-\beta d_2,\beta^2,a_2-\beta b_2,-\beta(b_2+d_2),a_1-\beta b_1,-\beta b_1,c_2,a_2+c_2,c_1,a_1+c_1,-\lambda_2/\beta)\Big] \label{eqS:Ifun}
\end{split}
\end{equation}
with $a_1 \dots d_2$ defined in \eqref{eqS:46}, $\beta$ defined in \eqref{eqS:45}, $p_{ij}$, $P_i$ defined in \eqref{eqS:44} and $\lambda_{1,2}$ solutions of \eqref{lam12}. Note that  parameters $\alpha$ and $\beta$ are solutions of quadratic equations \eqref{eqS:32} and \eqref{eqS:45} that in general have two real solutions each. The value of the four point function $\bar{D}_0(v, k_1,k_2,k_3,\Delta,M_1,M_2)$ does not depend on which of the solutions are chosen in the evaluation of \eqref{eqS:Tildefun},\eqref{eqS:Ifun}. This fact can be used as a useful test in the numerical implementation of the expressions given above. 

Now we take up the special  case of $p_1$ on the light-cone, i.e. $p_1^2=0$. From \eqref{eqS:32} it follows that  $\alpha=m_1^2/(m_1^2-m_2^2)$. If $m_1^2\ne m_2^2$ then $\alpha$ is finite and the calculation proceeds as before,  \eqref{eqS:10}-\eqref{eqS:8}. For equal  masses $m_1$ and $m_2$,  we evaluate the integrals by first taking $m_1^2\ne m_2^2$ and then performing the limit $m_1^2\to m_2^2$ and thus $|\alpha|\to \infty$. The external momenta in the last propagators of \eqref{eqS:5} are both equal to $l=\alpha p_1$. Thus the last external momentum in  $\tilde{D}_0$ of \eqref{eqS:6} is going to be $k_3=l=\alpha p_1$ for both of the integrals in \eqref{eqS:5}.  In the limit  $\alpha\to \infty$ the following leading order  values \eqref{eqS:46}  are obtained: 
$a_1\to 2 \alpha p_1\negthincdot(\beta v-  p_2)$,
$b_1\to 2 \alpha v\negthincdot p_1$,
$c_1\to 2 \alpha p_1\negthincdot p_2$
, where we have used also the fact that $p_1^2=0$. The other coefficients $a_2$, $b_2$, $c_2$, $d_2$ do not depend on $\alpha$. The first term in the denominator of the integrand in \eqref{eqS:7} is then  proportional to $\alpha$, while the second term in the denominator does not depend on $\alpha$ at all.  The $\alpha $ in the denominator  cancels against the $\alpha$ in the numerator of \eqref{eqS:5}. For the case of equal masses $m_1^2=m_2^2$ and $p_1^2=0$, the solution is then  the same as for the case  $m_1^2 \ne m_2^2$ except that (i) one has to replace $(1-\alpha)$ and $\alpha$ in \eqref{eqS:5} with $-1$ and $1$ respectively, and that (ii)  $a_1$, $b_1$, $c_1$ in \eqref{eqS:46}, \eqref{eqS:8} have to be replaced by their limiting values (divided by $\alpha$)
\begin{equation}
\begin{split}
a_1&\to a_1^l= 2 p_1\negthincdot (\beta v-p_2),\\
b_1&\to b_1^l= 2  v\negthincdot p_1,\\
c_1&\to  c_1^l=2  p_1\negthincdot p_2- i \delta , \label{eqS:34}
\end{split}
\end{equation}
where $\beta$ is the solution to \eqref{eqS:45} (with $P_i$ defined in \eqref{eqS:44}), and is different for the two integrals in \eqref{eqS:5}. In the limiting value of $c_1$ coefficient given in \eqref{eqS:34} an additional $- i\delta$ prescription has been added. As will be shown in the next paragraph, this  does not have any effect on the value of the  four-point function. It does  make possible, however,  to express the integrals in \eqref{eqS:8} in terms of the functions $I_1$ and $I_2$ as in \eqref{eqS:Ifun}.

It is easy to see, that the limiting procedure as explained above does lead to an unambiguous result. One might in principle worry that limits $m_1^2\to m_2^2$ taken from above and below, corresponding to the  limits $\alpha\to \infty$ and $\alpha\to -\infty$ respectively, would lead to different results. The question is most conveniently settled if the $\tilde{D}_0$ functions in Eq. \eqref{eqS:Tildefun} are replaced by the expressions given in Eq. \eqref{eqS:53}. Once the limit $m_1\to m_2$ is taken, the first factors of the integrands have the same limiting value. For $\alpha$ large, thus the leading term is
\begin{equation}
\begin{split}
\bar{D}_0\to -\alpha \int_0^\infty \negtwo 2 d\lambda \int_0^1 dy &[-2 \alpha p_1\negcdot p_2 y +2 \alpha v\negcdot p_1\lambda +2 \alpha p_1\negcdot p_2 - i\delta]^{-1} \times\\
& [(p_1-p_2)^2y^2+\lambda^2+2 v\cdot (p_1-p_2) y\lambda +\\
& \qquad +P_2\lambda +(-(p_1-p_{2})^2+m_2^2-m_3^2)y+m_3^2-i\delta]^{-1}\\
+\alpha \int_0^\infty \negtwo 2 d\lambda \int_0^1 dy &[-2 \alpha p_1\negcdot p_2 y +2 \alpha v\negcdot p_1\lambda +2 \alpha p_1\negcdot p_2 - i\delta]^{-1} \times\\
& [p_2^2y^2+\lambda^2-2 v\cdot p_2 y\lambda +\\
& \qquad +P_2\lambda +(-p_{2}^2+m_1^2-m_3^2)y+m_3^2-i\delta]^{-1}, \label{limit}
\end{split}
\end{equation}
with $P_2=2 (v\negthincdot p_2+\Delta)$. After collecting  the two integrands in \eqref{limit} the first factor in the integrands cancels and one finds 
\begin{equation}
\begin{split}
\bar{D}_0\to - \int_0^\infty \negtwo 2 d\lambda \int_0^1 y dy &[(p_1-p_2)^2y^2+\lambda^2+2 v\cdot (p_1-p_2) y\lambda +\\
& \qquad +P_2\lambda +(-(p_1-p_{2})^2+m_2^2-m_3^2)y+m_3^2-i\delta]^{-1}\times \\
& [p_2^2y^2+\lambda^2-2 v\cdot p_2 y\lambda +\\
& \qquad +P_2\lambda +(-p_{2}^2+m_1^2-m_3^2)y+m_3^2-i\delta]^{-1},
\end{split}
\end{equation}
This result exhibits clearly the fact that (i) the limit $m_1^2\to m_2^2$ is independent of whether it is taken from above or below and (ii) the limit is independent of the size (or even the sign) of the infinitesimal parameter in the first terms of the integrands in \eqref{limit}.

When the  momenta $p_1$, $p_2$ satisfy $(m_1-m_{2,3})^2<p_{1,2}^2< (m_{1}+ m_{2,3})^2$, rendering a complex  $\alpha$, the procedure outlined above in \eqref{eqS:10}-\eqref{eqS:Tildefun} cannot be applied directly. Starting from \eqref{eqS:10}, we then use the propagator identity \eqref{eqS:9} on the last two propagators in \eqref{eqS:10}, where we set $\alpha$ such that $l^2=(p_1+\alpha(p_2-p_1))^2=0$. This has a real solution for $\alpha$ since  $p_1$ and $p_2$  are timelike as has been assumed at the beginning of this paragraph. Changing the notation slightly we then have for the scalar four point function (omitting the $- i\delta$ prescription in the notation)
\begin{equation}
\begin{split}
\frac{i}{(2\pi)^{4}} &\int d^{4} q\frac{\alpha'}{(v\negcdot q -\Delta)[q^2-m_1^2][(q+p_1)^2-m_2^2][(q+l)^2-M^2]}\; +\\
\frac{i }{(2\pi)^{4}} &\int d^{4} q\frac{1-\alpha'}{(v\negcdot q -\Delta)[q^2-m_1^2][(q+p_2)^2-m_3^2][(q+l)^2-M^2]}, \label{eqS:11}
\end{split}
\end{equation}
with $\alpha' $ the (real) solution of 
\begin{equation}
(p_1+\alpha'(p_2-p_1))^2=0 \label{eqS:quadalphaPr}
\end{equation}
 and 
\begin{subequations}
\begin{align}
l&=p_1+\alpha'(p_2-p_1),\\
M^2&=(1-\alpha')m_2^2+\alpha' m_3^2-\alpha'(1-\alpha') (p_2-p_1)^2 .
\end{align}
\end{subequations}
The integrals in \eqref{eqS:11} can now be solved using the procedure outlined above \eqref{eqS:10}-\eqref{eqS:34}, once we permute the last two propagators with $l$ taking the role of $p_1$ in \eqref{eqS:10}. Note also, that $\alpha'$ solves quadratic equation \eqref{eqS:quadalphaPr} that in general has two solutions. The final result for the four point function $\bar{D}_0$ does not depend on which of the two solutions is taken in \eqref{eqS:11}. This fact can be  exploited in the numerical implementation as a useful check.

	The four point function has already been calculated before for a special case of $m_1=m_2=m_3$, $p_1^2=p_2^2=0$ and $p_1^\mu-p_2^\mu=Mv^\mu$ (see Eq. (A11) of  \cite{Fajfer:2001ad}). It has been checked numerically that the two solutions, the one given in this paper and the solution of \cite{Fajfer:2001ad}, agree for this special case. A number of other numerical tests have been performed. The direct numerical integration of \eqref{eqS:52} and the  evaluation of analytical result given in this paper have been found to agree numerically. It has been also checked  that the results do not depend on which of the  two solutions for $\alpha$, $\beta$ or $\alpha'$ is taken. The solution for the four-point function calculated in this paper has also been checked numerically to have the  branch cuts as required by analyticity and unitarity.

\section{Conclusions}
 In this article the general expressions for  the dimensionally regularized scalar  two-point, three-point and four-point functions with one heavy quark propagator were  derived. The calculation proceeds along similar lines as in the case of the one-loop scalar functions of the conventional perturbation theory. There are, however, also a number of differences that result from the appearance of the heavy quark propagator. Since this is linear in the integration variable, one is forced to use a modified Feynman parameterization \eqref{eqS:24}. In the calculation of the four-point function this prevents one to use the elegant formalism of \cite{'tHooft:1978xw},\cite{Denner:qq}. A repeated use of the propagator identity \eqref{eqS:9} then leads to the final expression for the four-point function. The simplification in the calculation is, that the heavy quark velocity $v^\mu$ is a timelike vector, which than leads to real solutions of the quadratic equations \eqref{eqS:49}, \eqref{eqS:45}. The complication on the other hand is the integration over infinite range that stems from the Feynman parameterization \eqref{eqS:24}.

The five-point as well as the  higher-point functions can be expressed in terms of the scalar functions given in this paper  using the standard procedure \cite{'tHooft:1978xw},\cite{vanNeerven:1983vr}. Consider for instance  the case of five-point scalar function. This is a function of four vectors, $v$ and $p_1,p_2 ,p_3$. The five-point function is first multiplied by $v_\mu \epsilon_{\alpha\beta\gamma\delta}$ and then antisymmetrized in all five indices. The resulting tensor is zero, because there is no antisymmetrical tensor with five indices in four dimensions. Then the tensor is multiplied first with $p_{1\alpha}p_{2\beta}p_{3\gamma}q_\delta$ and finally with  $v^\mu \epsilon_{\alpha'\beta'\gamma'\delta'}p_1^{\alpha'}p_2^{\beta'}p_3^{\gamma'}q^{\delta'}$. Using the  decomposition of the  product of two Levi-Civita tensors in terms of the Kronecker delta functions and expressing  the scalar products  $q \negthincdot p_i$ in terms of the  propagators $((q+p_i)^2- m_{i+1}^2)$ and $(q^2-m_1^2)$, the five-point function can be expressed in terms of the four point functions. The tensor functions can also be expressed in terms of the scalar functions using the algebraic reduction \cite{Passarino:1978jh}.

In the numerical implementation of the expressions given in this article,  further care has to be taken regarding the numerical instabilities. Such numerical instabilities can for instance arise, if one of the solutions of the quadratic equation is  much smaller than its coefficients. There is also a possibility of a cancellation between the dilogarithmic functions,  when the values of the dilogarithms separately are much larger then their sum.  These difficulties can be dealt with  along the lines of Ref. \cite{vanOldenborgh:1989wn}.
\vskip0.5cm
{\bf Acknowledgments}
\vskip0.1cm
I wish to thank S. Fajfer and especially B. Bajc for helpful discussions and for carefully reading the manuscript. It is a pleasure to also thank the theory group at Technion for warm hospitality during my visit. This work was supported in part by the Ministry of Education, Science and Sport of the Republic of Slovenia.

\appendix
\section{Dilogarithm and hypergeometric function}
\label{app:A}
In this appendix we list some properties of the dilogarithm and the hypergeometric function used in the rest of the paper (for other properties consult e.g. \cite{'tHooft:1978xw}, \cite{Gradshteyn} ).

The dilogarithm or Spence function is defined as 
\begin{equation}
\Li(x)=-\int_0^1 dt \; \frac{\ln(1-x t)}{t} \label{eqS:38}.
\end{equation}
The cut for the logarithm  along the negative real axis translates into the  cut for the dilogarithm along the  positive real axis for $x>1$. For $x$ on the positive real axis, $x>1$, dilogarithm is calculated using the following prescription $\Li (x) \to \Li (x- i\epsilon)$.  Note as well that $\Li(0)=0$.

Useful identities valid also for the complex arguments (not equal to zero)  are
\begin{subequations}
\begin{align}
\Li(x)&=-\Li\Big(\frac{1}{x}\Big)-\frac{1}{6}\pi^2 -\frac{1}{2}\big[\ln(-x)- 2 \pi i \xi(x)\big]^2\label{eqS:19},\\
\Li(x)&=-\Li(1-x)+\frac{1}{6}\pi^2-\ln(x)\ln(1-x), \label{eqS:16}
\end{align}
\end{subequations}
where
\begin{equation}
\xi(x)=
\left\{
\begin{aligned}
\;&1\; ; \; & x\in (0,1),\\
\;&0\; ; \; & \text{ otherwise}.
\end{aligned}
\right.
\end{equation}

The hypergeometric function for complex argument  $|z|<1$ is defined in  terms of the series
\begin{equation}
{}_2F_1(\alpha,\beta;\gamma;z)=1+\sum_{n=0}^{\infty}\frac{\alpha \dots (\alpha+n)\beta\dots (\beta+n)}{\gamma\dots (\gamma+n) (n+1)!}\;z^{n+1}, \label{eqS:15}
\end{equation}
with $\gamma$ not equal to zero or negative integer. Note that the series terminates if $\alpha$ or $\beta$ are equal to negative integer or zero. If either of them is zero then
\begin{equation}
{}_2F_1(0,\beta;\gamma;z)={}_2F_1(\alpha,0;\gamma;z)=1.
\end{equation}
For $z$ outside the unit circle the values of the  hypergeometric function can be obtained through analytic continuation. We make a cut in the $z$ plane along the real axis from $z=1$ to $z=\infty$. Then the series \eqref{eqS:15} will yield, in the cut plain,  a single valued analytic continuation that can be obtained using the following identity (other similar transformation formulas can be found  in e.g. \cite{Gradshteyn})
\begin{equation}
\begin{split}\label{eqS:14}
{}_2F_1(\alpha,\beta;\gamma;z)=&\frac{\Gamma(\gamma)\Gamma(\beta-\alpha)}{\Gamma(\beta)\Gamma(\gamma-\alpha)} (- z)^{-\alpha} {}_2F_1(\alpha,\alpha+1-\gamma;\alpha+1-\beta;1/z)\\
+&\frac{\Gamma(\gamma)\Gamma(\alpha-\beta)}{\Gamma(\alpha)\Gamma(\gamma-\beta)}(- z)^{-\beta} {}_2F_1(\beta,\beta+1-\gamma;\beta+1-\alpha;1/z) .
\end{split}
\end{equation}

The integral representations of the  hypergeometric function include
\begin{subequations}
\label{eqS:12}
\begin{align}
\int_z^\infty \frac{x^{\mu-1}dx}{(1+\beta x)^\nu}&=\frac{z^{\mu-\nu}}{\beta^\nu (\nu-\mu)} \; {}_2F_1(\nu,\nu-\mu;\nu-\mu+1;-1/(\beta z)) , \label{eqS:22}\\
\int_0^z \frac{x^{\mu-1}dx}{(1+\beta x)^\nu}& = \frac{z^\mu}{\mu} \; {}_2F_1(\nu,\mu;1+\mu;-\beta z) , \label{eqS:23}
\end{align}
\end{subequations}
where  Eq. \eqref{eqS:22} is valid for $\Re \nu>\Re\mu$, while Eq. \eqref{eqS:23} is valid for the case, when $\arg(1+\beta z)<\pi$ and $\Re\mu>0$.

\section{Reduction to dilogarithms}
\label{app:B}
In this appendix we will express the integrals appearing in \eqref{eqS:21}, \eqref{eqS:8} in terms of the dilogarithms.
First we review the derivations given in \cite{'tHooft:1978xw}. Consider
\begin{equation}
\begin{split}
R(\lambda_1,\lambda_0)& =\int_0^1 d\lambda \frac{1}{\lambda -\lambda_0}[\ln(\lambda-\lambda_1)-\ln(\lambda_0-\lambda_1)]\\
&=\int_{-\lambda_1}^{1-\lambda_1} d\lambda \frac{1}{\lambda -\lambda_0 +\lambda_1}[\ln\lambda-\ln(\lambda_0-\lambda_1)] ,  \label{eqS:18}
\end{split}
\end{equation}
where $\lambda_{0,1}$ may be complex. The residue of the pole of the integrand is zero. The cut of the logarithm is along the negative real axis, so for $\lambda_1$ not real, the cut is  outside the triangle $0$, $-\lambda_1$, $1-\lambda_1$. The integration path can thus be deformed to (for $\lambda_1$ real, this statement is trivial)
\begin{equation*}
\int_{-\lambda_1}^{1-\lambda_1} d\lambda =\int_0^{1-\lambda_1}d\lambda-\int_0^{-\lambda_1} d\lambda .
\end{equation*}
Making the substitutions $\lambda=(1-\lambda_1)\lambda'$ and $\lambda=\lambda_1\lambda'$ we obtain
\begin{equation}
\begin{split}
R(\lambda_1, \lambda_0)&=\int_0^1 d\lambda\Big[\frac{d}{d\lambda}\ln\Big(1+\lambda \frac{1-\lambda_1}{\lambda_1-\lambda_0}\Big)\Big][\ln \lambda(1-\lambda_1)-\ln(\lambda_0-\lambda_1)]\\
&-\int_0^1 d\lambda\Big[\frac{d}{d\lambda}\ln\Big(1-\lambda \frac{\lambda_1}{\lambda_1-\lambda_0}\Big)\Big][\ln(-\lambda\lambda_1)-\ln(\lambda_0-\lambda_1)] . \label{eqS:35}
\end{split}
\end{equation}
Since $\lambda$ is positive real, none of the arguments of the logarithms crosses the negative real axis. After integration per partes 
\begin{equation}
\begin{split}
R(\lambda_1,\lambda_0)=\Li\left(\frac{\lambda_1-1}{\lambda_1-\lambda_0}\right) +&\ln\left(\frac{1-\lambda_0}{\lambda_1-\lambda_0}\right)\left[ \ln(1-\lambda_1)-\ln(\lambda_0-\lambda_1)\right]\\
-\Li \left(\frac{\lambda_1}{\lambda_1-\lambda_0}\right)-&\ln\left(\frac{-\lambda_0}{\lambda_1-\lambda_0}\right)\left[ \ln(-\lambda_1)-\ln(\lambda_0-\lambda_1)\right] .
\end{split}
\end{equation}
This can be further simplified  using \eqref{eqS:16} 
\begin{equation}
\begin{split}
R(\lambda_1,\lambda_0)=&\Li\Big(\frac{\lambda_0}{\lambda_0-\lambda_1}\Big) +\left[\eta\Big(-\lambda_1, \frac{1}{\lambda_0-\lambda_1}\Big)+ 2 \pi i \Re^{(-)}(\lambda_0-\lambda_1)\right] \ln\frac{\lambda_0}{\lambda_0-\lambda_1}\\
-&\Li\Big(\frac{\lambda_0-1}{\lambda_0-\lambda_1}\Big)-\left[\eta\Big(1-\lambda_1, \frac{1}{\lambda_0-\lambda_1}\Big)+2\pi i\Re^{(-)}(\lambda_0-\lambda_1)\right] \ln\frac{\lambda_0-1}{\lambda_0-\lambda_1}, \label{eqS:51}
\end{split}
\end{equation}
with $\eta$ defined in \eqref{eqS:50} and $\Re^{(-)}(x)$ defined in \eqref{eqS:47}. Note that this result differs slightly from the one in \cite{'tHooft:1978xw} as  it is defined also for the arguments  lying on the negative real axis. The extension to negative real arguments  was not necessary in \cite{'tHooft:1978xw} as then the $\lambda_0$ was always real. This is not the case in the calculation of the four point function with one heavy quark propagator presented in this paper, as the $\lambda_1$ and $\lambda_2$ in \eqref{eqS:8} can have nonzero imaginary parts. The momenta and the masses in the calculation can then be chosen such, that one of the arguments appearing in \eqref{eqS:51} can lie on the negative real axis.

Next we turn to the integral 
\begin{equation}
S_3(a,b,c,\lambda_0)=\int_0^1 d\lambda \frac{1}{\lambda-\lambda_0}[\ln(a \lambda^2+b\lambda+c)-\ln(a \lambda_0^2+b\lambda_0+c)], \label{eqS:28}
\end{equation}
with $a$ real, while $b$, $c$, $\lambda_0$ may be complex but such, that the imaginary part of the  argument of the logarithm does not change sign for $x\in [0,1]$ (also $\Im c \ne 0$). 

Let $\epsilon$ and $\delta$ be infinitesimal quantities that have the {\it opposite} sign from the imaginary part of first and second argument of the logarithm respectively. That is, the signs of the arguments are as given by $-i\epsilon$ and $-i \delta$. Then
\begin{equation}
\begin{split}
S_3=&\int_0^1 d\lambda \frac{1}{\lambda-\lambda_0}[\ln(\lambda-\lambda_1)(\lambda-\lambda_2)-\ln(\lambda_0-\lambda_1)(\lambda_0-\lambda_2)]\\
&-\eta\Big(a-i\epsilon,\frac{1}{a-i\delta}\Big)\ln\Big(\frac{\lambda_0-1}{\lambda_0}\Big),\label{eqS:S3}
\end{split}
\end{equation}
with $\lambda_{1,2}$ the solutions of $a\lambda^2+b\lambda+c=a(\lambda-\lambda_1)(\lambda-\lambda_2)$. Next we split up the logarithms, use the fact that the imaginary part of $(\lambda-\lambda_1)(\lambda-\lambda_2)$ has the same sign as the  imaginary part of $c/a$ and use the definitions of $R(\lambda_1,\lambda_0)$ \eqref{eqS:18} to get
\begin{equation}
\begin{split}
S_3(a,b,c,\lambda_0)=&R(\lambda_1,\lambda_0)+R(\lambda_2,\lambda_0)\\
&+\Big[\eta(-\lambda_1,-\lambda_2)-\eta(\lambda_0-\lambda_1,\lambda_0-\lambda_2)-\eta\Big(a-i \epsilon, \frac{1}{a-i\delta}\Big)\Big]\ln\frac{\lambda_0-1}{\lambda_0} , \label{eqS:29}
\end{split}
\end{equation}
with $\epsilon$ and $\delta$ defined before Eq. \eqref{eqS:S3}.

For future reference we also define 
\begin{equation}
S_2(b,c,\lambda_0)=\int_0^1 d\lambda \frac{1}{\lambda-\lambda_0} [\ln(b \lambda+c)-\ln(b \lambda_0+c)],
\end{equation}
with $b$ real and $c$, $\lambda_0$ possibly complex ($\Im c \ne 0$). Defining as above infinitesimal parameters $\epsilon'$ and $\delta'$ to have signs {\it opposite} to the imaginary parts of the first and the second argument of the logarithms respectively, we obtain
\begin{equation}
S_2(b,c,\lambda_0)=R(-c/b, \lambda_0)-\eta\Big(b-i\epsilon',\frac{1}{b-i\delta'}\Big)\ln\frac{\lambda_0-1}{\lambda_0}.
\end{equation}

Next we turn to the integrals appearing in the calculations of the three-point  and four-point functions with one heavy quark propagator. Consider first
\begin{equation}
\begin{split}
I_1(a_1,a_2,b_1,b_2,b_3,b_4,&c_1,c_2,c_3,c_4,\lambda_0)=\\
&\int_0^1d\lambda \frac{1}{\lambda-\lambda_0} \Big[\ln\frac{b_3 \lambda+c_3}{b_4\lambda+c_4}-\ln\frac{a_1 \lambda^2+b_1\lambda+c_1}{a_2\lambda^2+b_2\lambda +c_2}\Big], \label{eqS:31}
\end{split}
\end{equation}
with $a_{1,2}$ and $b_{1,\cdots,4}$ real, while $\lambda_0$, $c_{1\cdots 4}$ may be complex but such that $\Im c_1 \Im c_2>0$ and $\Im c_3\Im c_4>0$. Also, the coefficients are such, that for $\lambda=\lambda_0$  the two logarithms are equal, so that the residuum of the integrand is equal to zero. Such an integral appears in the calculation of the four-point scalar function \eqref{eqS:8}. 
To reduce the integral $I_1$ to the integrals $S_2$, $S_3$ we add and subtract the values of the logarithms at the pole. Since the numerators and the  denominators of the logarithms in \eqref{eqS:31} have  imaginary parts of the same sign, we can split the logarithms.  Additional $\eta$ terms appear, however,  when we split the logarithms with $\lambda$ set to $\lambda_0$. As the result we get
\begin{equation}
\begin{split}
I_1&=S_2(b_3,c_3,\lambda_0)-S_2(b_4,c_4,\lambda_0)-S_3(a_1,b_1,c_1,\lambda_0)+S_3(a_2,b_2,c_2,\lambda_0)\\
&+\Big[\eta\Big(a_1\lambda_0^2+b_1\lambda_0+c_1,\frac{1}{a_2\lambda_0^2+b_2\lambda_0+c_2}\Big)+2 \pi i \Re^{(-)}(a_2 \lambda_0^2+b_2 \lambda_0+c_2)\\
&\qquad -\eta\Big(b_3\lambda_0+c_3,\frac{1}{b_4\lambda_0+c_4}\Big) -2 \pi i \Re^{(-)}(b_4\lambda_0+c_4)\Big]\ln\frac{\lambda_0-1}{\lambda_0}. \label{eqS:26}
\end{split}
\end{equation}
with $\eta$ defined in \eqref{eqS:50} and  $\Re^{(-)}$ in \eqref{eqS:47}.

Next consider the integral 
\begin{equation}
\begin{split}
I_2(a_1,a_2,g_1,f_1,g_2,&f_2,\lambda_0)=\\
&\int_0^\infty \negtwo d\lambda\frac{1}{\lambda-\lambda_0} \left\{ \ln\frac{\lambda-a_1}{\lambda-a_2}-\ln\frac{\lambda^2+g_1\lambda+f_1}{\lambda^2+g_2\lambda+f_2}\right\}, \label{eqS:36}
\end{split}
\end{equation}
with $g_{1,2}$ real, while $\lambda_0$, $a_{1,2}$, $f_{1,2}$ may be complex with the restriction  $\Im a_1 \Im a_2>0$, $\Im f_1\Im f_2>0$. Then the logarithms can be split without introducing $\eta$ terms, independent of the value of $\lambda$ as long as this is real. Also the arguments of the logarithms in \eqref{eqS:36} are taken to be the same for $\lambda=\lambda_0$, so that the residuum of the integrand is zero. Such integrals appear in the calculation of the three-point function \eqref{eqS:21} and in the calculation of the four-point function \eqref{eqS:8}.    We rewrite the integral \eqref{eqS:36} as
\begin{equation}
I_2=\int_0^\infty \negtwo d\lambda\frac{1}{\lambda-\lambda_0} \Big\{ \ln\frac{\lambda-a_1}{\lambda-a_2}-\ln\frac{(\lambda-b_1)(\lambda-b_2)}{(\lambda-c_1)(\lambda-c_2)}\Big\} , \label{eqS:30}
\end{equation}
with
\begin{equation}
\begin{split}
\lambda^2+g_1\lambda+f_1&=(\lambda-b_1)(\lambda-b_2),\\
\lambda^2+g_2\lambda+f_2&=(\lambda-c_1)(\lambda-c_2), \label{eqS:48}
\end{split}
\end{equation}
where $\Im b_1\Im b_2<0$, $\Im c_1\Im c_2<0$, $\Im(b_1b_2)\Im(c_1c_2)>0$ as can be seen from the constraints on $g_{1,2}$, $f_{1,2}$. Then the logarithms can be split up in the sum of the logarithms with arguments linear in $\lambda$.  

To the integral \eqref{eqS:30} we add logarithms with $\lambda$ set to $\lambda_0$ and then split the logarithms
\begin{equation}
\begin{split}
0&=\ln\frac{\lambda_0-a_1}{\lambda_0-a_2}-\ln\frac{(\lambda_0-b_1)(\lambda_0-b_2)}{(\lambda_0-c_1)(\lambda_0-c_2)}\\
&=\sum_i \rho(\kappa_i) \ln(\lambda_0-\kappa_i) -\eta' , \label{eqS:39}
\end{split}
\end{equation}
where $\kappa_i$ are the coefficients $a_{1,2}$, $b_{1,2}$, $c_{1,2}$ with $\rho(\kappa_i)=1$ for $a_1,c_{1,2}$ and $\rho(\kappa_i)=-1$ for $a_2,b_{1,2}$. There is also a sum of $\eta$ terms that we do not write out explicitly, but just denote by $\eta'$,  as it will be reabsorbed in the final result. Note also, that in the case of $\lambda_0 -\kappa_i$ real and negative  the logarithm  is  calculated using the prescription $\lambda_0-\kappa_i\to \lambda_0-\kappa_i + i \delta$, with $\delta$ a positive infinitesimal parameter (see also \eqref{eqS:55}-\eqref{eqS:54}). The integral is then
\begin{equation}
I_2=\sum_i \rho(\kappa_i) \int_0^\infty \negtwo d\lambda \frac{1}{\lambda-\lambda_0}[\ln(\lambda-\kappa_i)-\ln(\lambda_0-\kappa_i)] + \eta' \int_0^\infty \frac{d\lambda}{\lambda-\lambda_0} . \label{eqS:20}
\end{equation}
The separate integrals are divergent so they have to be regulated. We use the cutoff $M$ that is sent to infinity  at the end of the calculation. Note also, that there is no problem with the pole in the last term even if $\lambda_0$ is real, as then $\eta'$ is zero. 

The regulated integrals are then
\begin{equation}
\int_0^M d \lambda \frac{1}{\lambda-\lambda_0} [\ln(\lambda-\kappa_i)-\ln(\lambda_0-\kappa_i)].
\end{equation}
Let us from here on first assume, that $\lambda_0-\kappa_i$ is not negative real. Changing the variable $\lambda=M\lambda'$ and using the calculation of $R(\lambda_1,\lambda_0)$  \eqref{eqS:18}, \eqref{eqS:51} we get
\begin{equation}
\begin{split}
\int_0^1 &d\lambda \frac{1}{\lambda-\frac{\lambda_0}{M}} \left[\ln\left(\lambda-\frac{\kappa_i}{M}\right)-\ln\left(\frac{\lambda_0}{M}-\frac{\kappa_i}{M}\right)\right]=  \\
&\Li\frac{\lambda_0}{\lambda_0-\kappa_i}-\Li \frac{\lambda_0-M}{\lambda_0-\kappa_i}+\eta\Big(-\kappa_i, \frac{1}{\lambda_0-\kappa_i}\Big) \ln \frac{\lambda_0}{\lambda_0-\kappa_i} -\eta\Big(1-\frac{\kappa_i}{M}, \frac{M}{\lambda_0-\kappa_i}\Big) \ln\frac{\lambda_0-M}{\lambda_0-\kappa_i}.
\end{split}
\end{equation}
For $M$ big enough the last term is zero. The $M$ dependent dilogarithm can be transformed using relation \eqref{eqS:19}
\begin{equation}
\Li\frac{-M}{\lambda_0-\kappa_i}=-\Li\frac{\lambda_0-\kappa_i}{-M} -\frac{1}{6}\pi^2 -\frac{1}{2} \ln^2\Big(\frac{M}{\lambda_0-\kappa_i}\Big).
\end{equation}
The argument of the dilogarithm on the right-hand side goes toward zero as $M\to \infty$, so that in that limit the dilogarithm vanishes. Next we split the logarithm in  the last term and write
\begin{equation}
\ln^2\Big(\frac{M}{\lambda_0-\kappa_i}\Big)=\ln^2 M-2 \ln M\ln (\lambda_0-\kappa_i)+\ln^2(\lambda_0-\kappa_i).
\end{equation}
The first term gives zero once summed over in \eqref{eqS:20}, while the second term cancels against the $\eta'$ term in \eqref{eqS:20}. Leaving the case of $\lambda_0-\kappa_i$ negative real to the reader,  the final result is
\begin{equation}
\begin{split}
I_2=\sum_i \rho(\kappa_i) \bigg\{ &\Li \frac{\lambda_0}{\lambda_0-\kappa_i} +\left[\eta\Big(-\kappa_i, \frac{1}{\lambda_0-\kappa_i}\Big)+ 2 \pi i \Re^{(-)}(\lambda_0-\kappa_i)\right] \ln\frac{\lambda_0}{\lambda_0-\kappa_i}\\
+&\frac{1}{2}\ln^2(\lambda_0-\kappa_i)-\ln(\lambda_0-\kappa_i)\ln(-\lambda_0)\bigg\},\label{eqS:27}
\end{split}
\end{equation}
with
\begin{equation}
\rho(\kappa_i)=\left\{
\begin{aligned}
\;&+1; & \kappa_i=a_1, c_{1,2},\\
\;& -1; & \kappa_i=a_2,b_{1,2}. \label{eqS:37}
\end{aligned}
\right.
\end{equation}
Note that this solution applies also for the case encountered in the calculation of the three point function \eqref{eqS:21}, when $a_1=a_2=0$. Then the terms containing  $a_{1,2}$ cancel each other, so they can be dropped altogether for the case of Eq. \eqref{eqS:21}.
 
There is one more point  worth mentioning regarding the expression \eqref{eqS:27}. One might think that problems could arise for $\lambda_0-\kappa_i$ negative real or $\lambda_0/(\lambda_0-\kappa_i)$ real as then one has to deal with the cuts in the logarithm and the dilogarithmic function\footnote{Note that there exists such a combination of parameters $v$, $p_{1,2}$, $\Delta$ and $m_{1,2,3}$ in \eqref{eqS:10} that $\lambda_0-\kappa_i$ in \eqref{eqS:27} is negative real for some $i$, as can be seen from definition of $a_1,\dots,d_2$ \eqref{eqS:46}, definition of $\lambda_{1,2}$ \eqref{lam12} and the expression for the four-point function \eqref{eqS:Ifun}.}. We use the prescription for the  arguments lying exactly on the cuts of the functions as described before Eq. \eqref{eqS:55} and after Eq. \eqref{eqS:38}. One could  as well use a different prescription, with infinitesimal parameter $\epsilon$ in  \eqref{eqS:55}, \eqref{eqS:38} taken to be negative, and with appropriately adjusted $\eta$ and $\Re^{(-)}$ functions. It has been checked numerically that the result \eqref{eqS:27} does not depend on which prescription is used. Thus the result \eqref{eqS:27} is valid for any complex $\lambda_0$, $\kappa_i$ independent of the prescription used for the arguments lying on the cut.

For the special case of $\lambda_0$ real the result \eqref{eqS:27} simplifies considerably. The $\eta$ term is then zero. Also the last term in \eqref{eqS:27}, that arises from the $\eta'$ term in \eqref{eqS:20}, then sums up to zero. For $\lambda_0$ real we have
\begin{equation}
I_2=\sum_i \rho(\kappa_i) \Big[ \Li \frac{\lambda_0}{\lambda_0-\kappa_i}+ \frac{1}{2}\ln^2(\lambda_0-\kappa_i)\Big], \label{eqS:41}
\end{equation}
with $\kappa_i$ and $\rho(\kappa_i)$ as in \eqref{eqS:37}. 

Of special interest  is the case of $\lambda_0$ almost real, i.e. $\lambda_0=\lambda_0^{re}+i \delta' $, where $\lambda_0^{re}$ is the real part of $\lambda_0$ and $\delta'$ an infinitesimal (not necessarily positive) parameter. One can of course still use the solution \eqref{eqS:27}. The problem is, however,  that for both $\kappa_i$ and $\lambda_0$ almost real one has to keep track of the relative sizes of the infinitesimal imaginary parts. This complication can be avoided by the following procedure. First we set $\lambda_0$ in the second line of \eqref{eqS:39} equal to its real part. By doing this, the arguments of the logarithms can cross the negative real axis, which is compensated by a new sum of $\eta$ functions, $\eta''$. Then  instead of \eqref{eqS:20} we have
\begin{equation}
\sum_i \rho(\kappa_i) \int_0^\infty \negtwo d\lambda \frac{1}{\lambda-\lambda_0^{re}- i\delta'}[\ln(\lambda-\kappa_i)-\ln(\lambda_0^{re}-\kappa_i)] +\eta'' \int_0^\infty \frac{d\lambda}{\lambda-\lambda_0}.
\end{equation}
In the first integral $\delta'$ can be safely put equal  to zero as the resulting integrand has vanishing residuum, with the logarithm in the  numerator being an analytic function in some neighbourhood of the pole (since $\Im(\kappa_i)\ne0)$. The integral thus does not depend on how we avoid the pole (i.e. $\delta'$ can be positive, negative or zero). In the second integral one has to keep the imaginary part of $\lambda_0$.

The final result for almost real $\lambda_0=\lambda_0^{re}+i\delta'$  is then
\begin{equation}
I_2=\sum_i \rho(\kappa_i) \Big[ \Li \frac{\lambda_0^{re}}{\lambda_0^{re}-\kappa_i} 
+\frac{1}{2}\ln^2(\lambda_0^{re}-\kappa_i)-\ln(\lambda_0^{re}-\kappa_i)\ln(-\lambda_0)\Big], \label{eqS:40}
\end{equation}
where in the last logarithm $\lambda_0$ is kept together with its  infinitesimal imaginary part.

\end{document}